\documentclass[pdflatex,sn-mathphys,iicol,dvipsnames]{sn-jnl}% Math and Physical Sciences Reference Style
\usepackage[utf8]{inputenc}
\usepackage[T1]{fontenc}
\usepackage{xcolor}
\usepackage{bm}
\usepackage{xspace}
\usepackage{siunitx}

\newcommand{\OpenFOAM}{OpenFOAM\textsuperscript{\tiny\textregistered}\xspace}
\newcommand{\Wi}{$\mathrm{Wi}$\xspace}

\newcommand{\Karman}{K\'arm\'an\xspace}
\newcommand{\mean}[1]{\overline{#1}}

\jyear{2024}%

\begin{document}

\title[Mixing with elastic turbulence]{Mixing in viscoelastic fluids using elastic turbulence}


\author*[1]{\fnm{Reinier} \spfx{van} \sur{Buel}}\email{r.vanbuel@tu-berlin.de}

\author[1]{\fnm{Holger} \sur{Stark}}\email{holger.stark@tu-berlin.de}

\affil*[1]{\orgdiv{Institute of Theoretical Physics}, \orgname{Technische Universit{\"a}t Berlin}, \orgaddress{\street{Hardenbergstrasse 36}, \postcode{10623}, \state{Berlin}, \country{Germany}}}

\abstract{We investigate the influence of elastic turbulence on mixing 
{of a scalar concentration field}
within a viscoelastic fluid in a two-dimensional Taylor-Couette geometry using numerical solutions of the Oldroyd-B model. 
The flow state is determined through the secondary-flow order parameter indicating 
{the regime of elastic turbulence.}
When {starting in the turbulent state and subsequently} lowering the Weissenberg number, a weakly-chaotic flow occurs below $\text{Wi}_c$.
Advection in both {the turbulent and weakly-chaotic} flow states induces mixing, which we illustrate by the time evolution of the standard deviation of the solute concentration from the uniform distribution. In particular, in the elastic turbulent state mixing is strong and we quantify it by the mixing rate, the mixing time, and the mixing efficiency. All three quantities follow scaling laws.
Importantly, we show that the order parameter is strongly correlated to the mixing rate and hence is also a good indication of mixing within the fluid.}

%\keywords{keyword1, Keyword2, Keyword3, Keyword4}



\maketitle

\section{Introduction}\label{sec:intro}
Mixing in Newtonian fluids at low Reynolds numbers, where flow is laminar, is a slow process governed by diffusion. 
Remarkably, after adding some high-molecular-weight polymers,
the fluid becomes viscoelastic and can exhibit elastic turbulence even though inertia is negligible \cite{groisman2000,groisman2001efficient,groisman2004elastic}. 
This enables efficient mixing of fluids at very small scales \cite{burghelea2004chaotic,burghelea2004mixing,thomases2009transition,thomases2011stokesian,kumar1996chaotic,niederkorn1993mixing,arratia2006elastic,burghelea2007elastic,jun2017polymer,vanBuel2018elastic}, which is an important application for lab-on-a-chip devices\ \cite{squires2005microfluidics}. 

In flows with curvilinear streamlines, where strong shear deformations occur, the resulting elastic stresses can exceed the viscous stresses.
Thereby, they cause a secondary flow, which modifies the laminar base flow \cite{balkovsky2001turbulence}. 
Thus, these so-called hoop stresses acting perpendicular to the laminar flow drive an elastic instability in curved geometries \cite{larson1990purely}.
For example, in Taylor-Couette flows the polymers are mainly aligned along the base flow in azimuthal direction. The hoop stresses generate radial flow, which amplifies stretching of the polymers.
This, in turn, enhances the hoop stresses and the radial flows become even stronger \cite{larson1990purely}. 
The resulting elastic instability is governed by the Weissenberg number, the product of the fluid relaxation time and the shear rate. 
It has been thoroughly studied in experiments \cite{larson1990purely,mckinley1991observations,byars1994spiral,ducloue2019secondary,pakdel1996elastic}.
{In theory, the type of instability in the three-dimensional Taylor-Couette geometry has also been investigated. For example, using linear \cite{avgousti1993viscoelastic,renardy1996hopf} 
and nonlinear \cite{avgousti1993viscoelastic,sureshkumar1994non,renardy1996hopf} stability analysis, it was shown that the most unstable modes are three-dimensional 
and time-dependent. In particular, using the upper-convected Maxwell model, Ref.\ \cite{sureshkumar1994non} showed that the two most 
unstable modes are non-axisymmetric ribbon and spiral flow patterns, where at least one of them bifurcates subcritically for narrow gaps, while
at sufficiently wide gaps they both exhibit a supercritical instability. Indeed, in recent numerical simulations in wide gaps, both
flow patterns were observed \cite{song2022direct,song2023self}.}

{Ultimately,}
 at high Weissenberg numbers, the flow of viscoelastic fluids in geometries with curved streamlines becomes turbulent \cite{groisman2000,groisman2001efficient,groisman2004elastic,burghelea2007elastic,jun2017polymer,varshney2018drag,jun2009power,sousa2018purely,steinbergscaling}. The flow state was denoted elastic turbulence \cite{groisman2000}, due to the similarity to inertial turbulence, observed in Newtonian fluids at high Reynolds numbers.
The important characteristics of elastic turbulence are increased velocity fluctuations, a characteristic power-law decay observed in the temporal and spatial velocity power spectra, and an increased flow resistance \cite{groisman2000,groisman2001stretching,groisman2004elastic}.
The observed power-law scaling in the velocity spectrum in numerical work \cite{berti2010elastic,vanBuel2018elastic,canossi2020elastic,steinberg2021elastic} is independent of the viscoelastic model employed and agrees well with the values for the exponents observed in experiments \cite{groisman2000,groisman2001efficient,groisman2004elastic,burghelea2007elastic,jun2017polymer,varshney2018drag,jun2009power,sousa2018purely,steinbergscaling}.
In recent numerical simulations of elastic turbulence in a three-dimensional Taylor-Couette geometry with a wide gap, non axisymmetric 
and time-dependent ribbon (standing waves) and spiral (traveling waves) flow patterns have been observed \cite{song2022direct,song2023self}.

{
Flows of viscoelastic fluids with straight streamlines also experience elastic turbulence at high Weissenberg 
and low Reynolds numbers. 
{A weakly}
non-linear stability analysis predicts mostly subcritical instabilities in channel flows \cite{morozov2005subcritical,buza2022weakly}. In contrast, 
when polymer-stress diffusion is explicitly taken into account, 
planar Couette flow is linearly unstable \cite{beneitez2022linear}. 
A subcritical instability can be triggered by finite-size perturbations in the velocity field in a microchannel \cite{pan2013nonlinear} and is even observed without any added perturbations to the velocity field \cite{li2022elastic}.
Additionally, in the unbounded two-dimensional periodic Kolmogorov flow numerical simulations performed at low Reynolds numbers \cite{berti2008two} show a transition towards time-periodic flows that ultimately become randomly fluctuating \cite{berti2010elastic}.
Furthermore, an important element in elastic turbulence is the presence of elastic waves \cite{fouxon2003spectra,berti2010elastic,varshney2019elastic,song2023self}. 
{They}
are essential in the self-sustaining cycle of the turbulent dynamics and significantly contribute to the 
{dominant} flow structures \cite{jha2021elastically,song2023self}.
}

Importantly, the enhanced velocity fluctuations in the flow states following the elastic instability and, in particular, in the elastic turbulent state increase the transport of heat and mass. In time-dependent flows confined between two cylinders at small Weissenberg numbers, the elastic instability has a large effect on the advection of passive tracers, where the observed rate of stretching of the fluid elements is exponential \cite{niederkorn1993mixing,kumar1996chaotic}.
Experiments on the enhancement of mixing due to elastic turbulence in a curvilinear channel exhibit an exponential decay in the variance of the tracer density {\cite{groisman2001efficient}}. These results are in very good agreement with the theoretical predictions for mixing in the Batchelor regime \cite{groisman2001efficient}.
Further experiments demonstrated the advection of a blob of a passive scalar by elastic turbulence in the von \Karman swirling flow \cite{groisman2004elastic,burghelea2007elastic,poole2012emulsification}.
Moreover, elastic turbulence increases heat transfer, which is shown in experiments performed in a serpentine channel \cite{yang2020experimental} and in simulations of a lid-driven cavity \cite{gupta2022influence}.
In this work, Gupta \emph{et al.} determine
the heat transfer rate inside the cavity and show that it doubles in the presence of elastic turbulence compared to Newtonian fluid flow.

The enhancement of mixing in viscoelastic fluids has also been investigated in extensional flows. 
Breaking the symmetry of the flow in a four-roll mill increases fluid mixing in the region outside of the dominant vortex with increasing \Wi \cite{thomases2011stokesian}.
Also, by cycling the dominant vortex through all four quadrants a more global mixer is obtained \cite{thomases2011stokesian}. 
In experiments with a cross-slot geometry the elastic instability in the extensional flow increases mixing in the outlet region, as well \cite{arratia2006elastic}. 
Finally, in simulations of a channel flow through a periodic array of cylinders, an elastically driven transition occurs, which enhances the mixing layer near the channel walls \cite{grilli2013transition}.

In this article, we investigate the mixing of a solute due to elastic turbulence in a viscoelastic fluid at low Reynolds number, such as employed in microfluidic flows.
We study the two-dimensional Taylor-Couette flow using numerical solutions of the Oldroyd-B model. 
In our earlier work \cite{vanBuel2018elastic,vanBuel2020active}, we have shown that the study in two dimensions already provides significant insights into elastic turbulence compared to three-dimensional simulations \cite{vanBuel2022characterizing}.
Extending this work, we classify the transition from laminar to elastic turbulent flow as subcritical and identify a weakly-chaotic-flow state upon lowering \Wi. We show that advection in both {the turbulent and weakly-chaotic} flow states induces mixing, which is stronger and proceeds exponentially within elastic turbulence. The corresponding mixing rate is strongly correlated with the
second-flow order parameter and we also introduce a mixing time and mixing efficiency.

The remainder of the article is structured as follows. 
First, we present our employed methods comprising the geometry, constitutive equation, the simulation method, and parameters in Sec.\ \ref{sec:methods}. Next, in Sec.\ \ref{sec:results} we show the main results. We quantify the weakly-chaotic and turbulent flow states using an order parameter in Sec.\ \ref{subsec.subcritical}. Mixing in both states and, in particular, with elastic turbulence is discussed in Sec.\ \ref{subsec:mixing}. Finally, our conclusions are presented in Sec.\ \ref{sec:conclusion}.

\begin{figure}[t!]
\begin{center}
        \includegraphics[width=.33\textwidth]{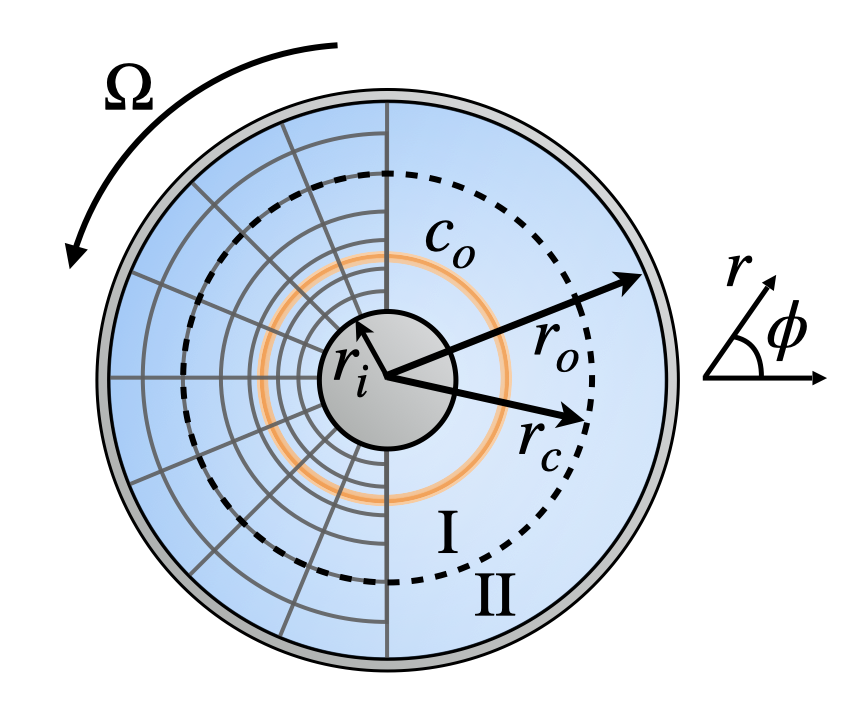}
\end{center}
\caption[Schematic of the two-dimensional Taylor-Couette geometry.]{
Schematic of the two-dimensional Taylor-Couette geometry. 
In the left half, the spokes-wheel
mesh used in \OpenFOAM with $N_r = 100$ cells in the radial direction and $N_{\phi} = 120$ cells in the angular direction 
is depicted. The radii of the inner and outer cylinders are $r_i = \SI{2.5}{\mu m}$ and $r_o = \SI{10}{\mu m}$, respectively.
The outer cylinder rotates with an angular velocity $\Omega = 2 \pi \SI{}{\per\second}$.
To investigate mixing, the area in between the cylinders is divided into two equal parts, an inner region denoted $\mathrm{I}$ with $r_i \leq r \leq r_c$ and an outer region denoted $\mathrm{II}$ with {$r_c < r \leq r_o$}, where $r_c=\sqrt{17/32}$. The initial distribution {$c_0$} of the concentration field {of a scalar,} used for mixing, is displayed in orange.
}
\label{fig:schematic}
\end{figure}

\section{Methods}\label{sec11}
\label{sec:methods}
In this section, we describe the geometry of the Taylor-Couette flow and the employed constitutive equation to model the viscoelastic fluid. Secondly, we discuss our simulation method and the simulation parameters used in this work.

\subsection{Geometry and model fluid}

We consider an incompressible viscoelastic fluid in a two-dimensional Taylor-Couette geometry consisting of two concentric cylinders with $r_o$ the radius of the outer cylinder and $r_i$ the radius of the inner cylinder. The outer cylinder is rotated counter-clockwise with a constant angular velocity $\Omega$ and the inner cylinder is fixed. 
A schematic of our set-up can be seen in fig.~\ref{fig:schematic}.
The dynamics of the flow field $\bm{u}(\bm{r},t)$ as a function of the position $\bm{r}$ and time $t$ can be described with a generalised Navier-Stokes equation:
\begin{equation}
\label{eq:NS} 
\rho \Big(
\frac{\partial \bm{u}}{\partial t} + \bm{u} \cdot \nabla \bm{u}  \Big)
= - \nabla p + \eta_s \nabla^2 \bm{u} +  {\nabla \bm{\tau}} \, .
\end{equation}
Here, $\rho$ is the density of the solvent, $p$ the pressure, $\eta_s$ the solvent shear viscosity, and ${\nabla \bm{\tau}}$ denotes the divergence of the stress tensor $\bm \tau$.
We use the characteristic length $r_o$, the characteristic time $\Omega^{-1}$, 
and the characteristic velocity $r_o\Omega$ to rescale length, time, and velocity, respectively. 
Therefore, the Reynolds number becomes \mbox{$\mathrm{Re} = \rho \Omega r_o^2 / \eta_s$}.

The additional viscoelastic stresses due to the dissolved polymers are described by the polymeric stress tensor $\bm{\tau}$, for which we choose the constitutive relation of the
Oldroyd-B model \cite{oldroyd1958non,oldroyd1950formulation}:
\begin{equation}
\label{eq:Oldroyd}
\bm{\tau} +\lambda \overset{\nabla}{\bm{\tau}} = 
\eta_p \left[ \nabla \otimes \bm{u} + (\nabla \otimes \bm{u})^\mathrm{T} \right] \, ,
\end{equation}
where $\eta_p$ is the polymeric shear viscosity and
$\lambda$ the characteristic relaxation time of the dissolved polymers. Finally, 
$\overset{\nabla}{\bm{\tau}}$ is the 
upper convective derivative of the stress tensor defined as
\begin{equation}
\overset{\nabla}{\bm{\tau}} = \frac{\partial  \bm{\tau}}{\partial t} + \bm{u}\cdot \nabla\bm{\tau} 
- (\nabla \otimes \bm{u})^\mathrm{T} \bm{\tau} - \bm{\tau}  ( \nabla \otimes \bm{u})  \, .
\end{equation}
Equations (\ref{eq:NS}) and (\ref{eq:Oldroyd}) can be written in dimensionless form with three relevant parameters.
Besides the Reynolds number $\mathrm{Re}$, one has the ratio $\beta = \eta_p/\eta_s$ of the polymeric shear viscosity to the solvent shear viscosity and the Weissenberg number $\mathrm{Wi} =  \lambda \dot{\gamma}$, where $\dot{\gamma}$ is the shear rate.
For the Taylor-Couette geometry the characteristic shear rate is the angular velocity of the outer cylinder, $\dot{\gamma} \sim \Omega$, and we have $\mathrm{Wi} \simeq \lambda \Omega$.

Considering an axisymmetric flow, which corresponds to our 2D geometry, a steady and laminar solution of Eqs.~(\ref{eq:NS}) and (\ref{eq:Oldroyd}) has been found \cite{larson1990purely}.
It agrees with the Taylor-Couette flow of a Newtonian fluid at low Re in the same geometry.
The solution, which we denote as the base flow $\bm{u}^0$ {and base elastic stress
{tensor}
 $\bm{\tau}^0$ }, is given by the simple shearing flow
{
\begin{align}
 & u_r^0 = 0  \, ;   u_\phi^0 = A r + B r^{-1} \, ,
\label{eq.TCflow}
\end{align}
{the stress tensor components}
\begin{align}
&\tau_{rr}^0 = 0 \, , \\
&\tau_{r\phi}^0 = {\tau_{\phi r}^0 =}
 -2 \eta_p B r^{-2} \, , \\
&\tau_{\phi\phi}^0 = 8 \eta_p \lambda B^2 r^{-4} \, ,
\end{align}
{and}
}
with
\begin{equation}
 A = \frac{r^2_o }{r^2_o-r^2_i}\Omega  \, ; \qquad B = -r^2_i A \Omega \, .
\end{equation}

To model the mixing of a solute within the fluid, we implement the advection-diffusion equation for the concentration field $c(\mathbf{r},t)$, 
\begin{align}\label{eq:AdvectionDiffusion}
 \frac{\partial c}{\partial t} + \left(\mathbf{u} \cdot \nabla\right) c = D\nabla^2 c \, ,
\end{align}
where $D$ is the molecular diffusion coefficient of the passive scalar. In the following, we neglect it by setting $D=0$.
We solve the advection-diffusion equation together with the governing Eqs.~(\ref{eq:NS}) and (\ref{eq:Oldroyd}). 
To solve Eq.\ (\ref{eq:AdvectionDiffusion}), we adopt the bounded \textit{van Leer} scheme for discretising the equation, 
{to avoid}
unphysical negative concentration values and 
{lessen the effect of}
non-physical numerical diffusion \cite{versteeg2007}.

\subsection{Simulation method and parameters}
\label{subsec:methods}
All our numerical results are obtained with the open-source finite-volume solver \OpenFOAM{} for computational fluid dynamics simulations 
on polyhedral grids. We give all parameters in SI units, as required by \OpenFOAM{}. We adopt the specialised solver for viscoelastic 
flows  RheoTool, develop by Favero et al. \cite{favero2010}, {in which the advection-diffusion equation 
{for the concentration field $c$}
is already implemented.
{The solver}}
has been tested for accuracy in benchmark flows and has been shown to have second-order accuracy in space and time \cite{pimenta2017}.

Between the two coaxial cylinders of our 2D geometry, we use a grid with mesh refinement towards the inner cylinder, where velocity 
gradients become larger. In the left part of the geometry presented in Fig.\ \ref{fig:schematic}, we sketch the grid mesh, which 
resembles a spokes wheel with $N_r = 100$ cells in the radial direction and $N_{\phi} = 120$ cells in the angular direction. The mesh 
refinement is such that the ratio of the radial grid size at the inner cylinder to the one at the outer cylinder is 4. The time step of the 
simulation is $\delta t = 10^{-5}\mathrm{s}$, where the velocity, pressure, scalar concentration, and stress fields are extracted every 
5000 steps.  At the two bounding cylinders we choose the no-slip boundary condition for the velocity, zero gradient for the pressure field
{in the direction normal to the wall,}
and an extra\-po\-lated zero gradient {in the direction normal to the wall} for the polymeric stress field following Ref.~\cite{pimenta2017,vanBuel2018elastic}.
We use a biconjugate gradient solver combined with a diagonal incomplete LU preconditioner (DILUP-BiCG) to solve for the components of the polymeric stress tensor and a conjugate gradient solver coupled to a diagonal incomplete Cholesky preconditioner (DIC-PCG) to solve for the velocity and pressure fields \cite{pimenta2017}.

The discretised advection-diffusion equation for the solute particles is also solved using the biconjugate gradient solver combined with a diagonal incomplete LU preconditioner, with zero gradient boundary conditions at the two bounding cylinders.

The simulations for investigating mixing are initiated with the viscoelastic fluid in the elastic turbulent state.
The turbulent state is reached by starting with the fluid at rest at $\mathrm{Wi}= 27.6$, where pressure, flow, and stress fields are uniformly zero, and rotating the outer cylinder for $t = 250$ rotations. Then, the Weissenberg number is lowered to the desired value.
This approach is also used when determining the order parameter for decreasing $\mathrm{Wi}$ in Sec.\ \ref{subsec.subcritical}; for increasing $\mathrm{Wi}$ we just choose the specific value of $\mathrm{Wi}$ and proceed with the outlined protocol.
The following geometric parameters are chosen from the viewpoint of microfluidic settings in such a way as to set a low Reynolds 
number. We adjust the Weissenberg number by varying the polymeric relaxation time $\lambda$.
We set the polymeric shear viscosity to $\eta_p=0.0015 \, {\mathrm{kg\,m^{-1}\,s^{-1}}}$, the solvent shear viscosity to 
\mbox{$\eta_s=0.001  \, {\mathrm{kg\,m^{-1}s^{-1}}}$}, and the density to $\rho=1000 \,\mathrm{kg/m^3}$.
The ratio of the polymeric to the solvent viscosity is then $\beta={\eta_p}/{\eta_s}=1.5$. 
Thus, the Reynolds number becomes \mbox{$\mathrm{Re} = 0.628 \cdot 10^{-3}$}.
The fluid flow is simulated up to ${600}\mskip3mu\mathrm{s}$.

\section{Results}\label{sec:results}

\begin{figure}
    \centering
    \vspace*{-2pt}
    \includegraphics[width=.45\textwidth]{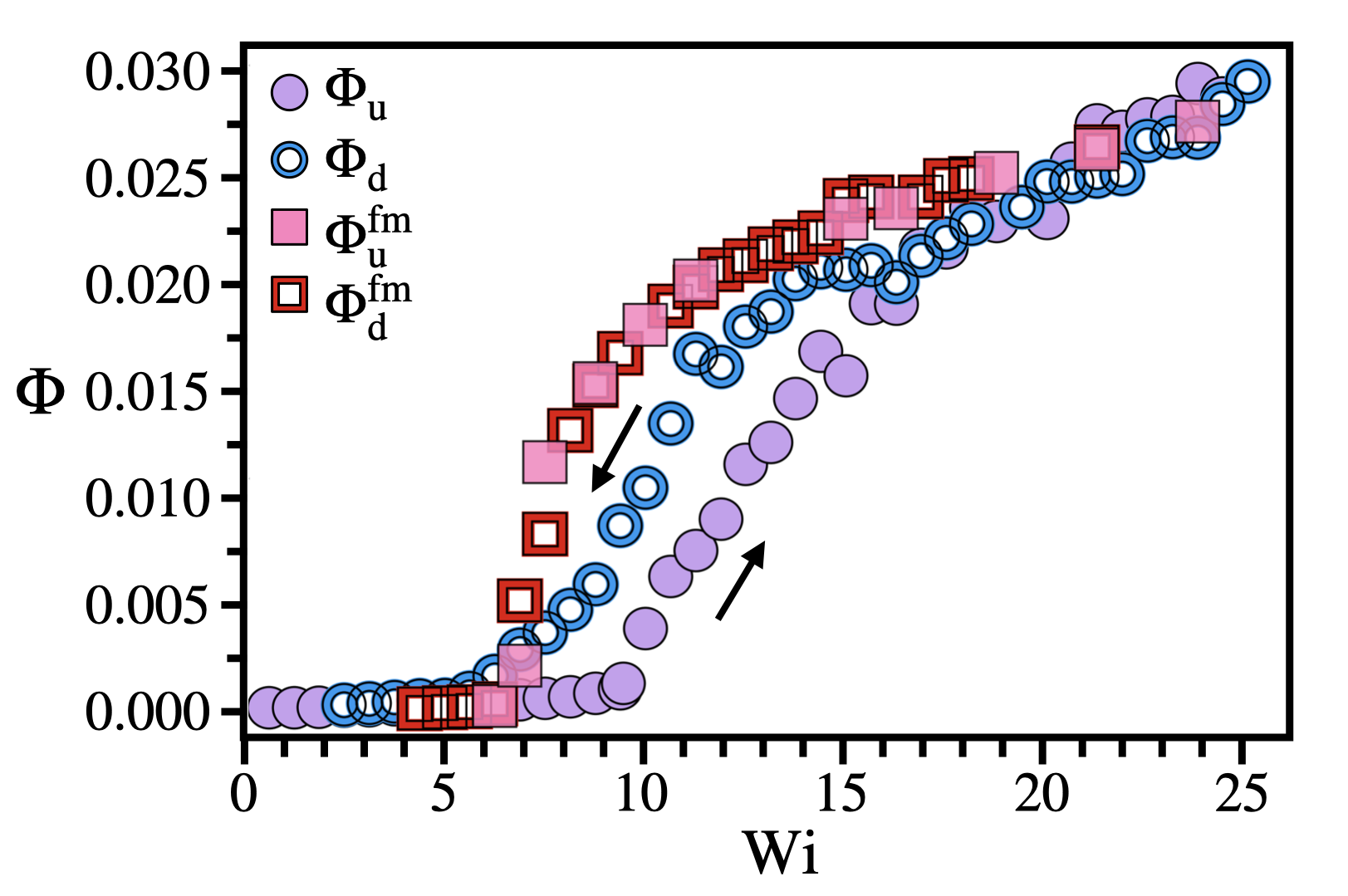}  
\caption{
{Order parameter $\Phi$ versus the Weissenberg number \Wi, initiated from rest  ($\Phi_\mathrm{u}$) or from the turbulent state ($\Phi_\mathrm{d}$) for the original mesh
with $N_r = 100 $ and $N_\phi = 120$. Hysteresis is observed between $\mathrm{Wi} \approx 5.5$ and $\mathrm{Wi} \approx 16$. The upward and downward arrow in the plot 
indicate an increase and decrease in \Wi compared to the initial state, respectively.
In the finer mesh, where $N_r = 150 $ and $N_\phi = 240$, hysteresis is not observed.}
}
\label{fig:OPhysteresis}
\end{figure}

First, 
{we 
revise our results on the nature of the transition presented in our earlier work \cite{vanBuel2018elastic}. 
For this,}
we characterise the different flow states using an order parameter. Then, we address mixing in a viscoelastic fluid by analysing the 
spreading of a passive scalar such as solutes in the different flow states. Finally, we introduce three measures to characterise the 
degree of mixing of the scalar field.

\subsection{Transition to elastic turbulence}
\label{subsec.subcritical}

As in our earlier work \cite{vanBuel2018elastic}, we take the secondary-flow strength
\begin{equation}
\sigma(t) \equiv\left. \sqrt{\left\langle [\bm{u}(\bm{r},t)-\bm{u}^0(\bm{r})]^2 \right\rangle}\right/u^0_{\mathrm{max}} \, ,
\label{eq.standard}
\end{equation}
as a measure for the velocity fluctuations of the flow field about the base flow $\bm{u^0}(\bm{r})$ and define an order parameter, $\Phi = \mean{\sigma}$, as its time average.
Here, $\langle \dots \rangle$ denotes a volume average performed over the radial and angular coordinate and $u^0_{\mathrm{max}}$ is the velocity of the outer cylinder.

The order parameter is plotted in Fig.~\ref{fig:OPhysteresis} as a function of the Weissenberg number. Two different initial conditions are used: one, where the fluid starts from rest ($\Phi_\mathrm{u}$), and one, where the fluid starts from the turbulent flow state at a high Weissenberg number, which then is subsequently reduced ($\Phi_\mathrm{d}$). For both initial conditions the flow is equilibrated during 100 rotations.
{This ensures that the secondary flow strength $\sigma(t)$ fluctuates about a steady value. Then, the}
order parameter is obtained from 
{$\sigma(t)$}
by averaging over another 400 - 500 rotations.
{Using the method of block averaging \cite{flyvbjerg1989error,frenkel2023understanding}, we checked that the statistical errors of the data points are not larger than the symbol size in 
Fig.\ \ref{fig:OPhysteresis}.}

In the case where the fluid starts at rest, a continuous transition from the laminar base flow to the elastic turbulent state is observed at a critical Weissenberg number $\mathrm{Wi_c}=9.95$. Analogously to our earlier work \cite{vanBuel2018elastic}, we determined the power spectra of the velocity fluctuations and identified the turbulent state from the steepness of the observed power-law decay (data not shown).
In the case where the fluid starts from the turbulent state and upon decreasing $\mathrm{Wi}$, the order parameter evolves continuously from the elastic turbulent flow state \mbox{($\Phi_\mathrm{d} > 0.007$)} to a weakly-chaotic state \mbox{($0< \Phi_\mathrm{d} < 0.007$)}, and to the laminar base flow \mbox{($\Phi_\mathrm{d} \approx 0$)}.
The observation of the weakly-chaotic flow state below $\mathrm{Wi_c}$ is similar
to our results in the von \Karman flow \cite{vanBuel2022characterizing}.

Most importantly, we find hysteretic behaviour in the order parameter, where $\Phi_\mathrm{d} > \Phi_\mathrm{u}$ between $\mathrm{Wi} \approx 5.5$ and $\mathrm{Wi} \approx 16$. 
{This would indicate}
that the nature of the transition from laminar to turbulent flow is subcritical, in contrast to our earlier findings in Ref.~\cite{vanBuel2018elastic}, where we 
{reported}
a supercritical transition
{since we}
did not check for hysteresis.
{However, as a surprising result, upon increasing the resolution of the mesh, where $N_r = 150 $ and $N_\phi = 240$, the transition shifts to lower Wi and the 
hysteretic behaviour disappears, see Fig.~\ref{fig:OPhysteresis}. The necessary computational resources do not allow us to currently investigate this further.
Therefore, at the moment, we cannot make any clear statement on the nature of the transition. We also notice that beyond $\mathrm{Wi}=10$, there is 
{little} difference 
between the branch $\Phi_\mathrm{d}$ and the branches $\Phi_\mathrm{u}^\mathrm{fm}$, $\Phi_\mathrm{d}^\mathrm{fm}$ of the finer mesh, in particular, beyond 
$\mathrm{Wi} = 20$ they fall on top of each other within the statistical errors.}

{So, to study mixing in viscoelastic fluids, {in a two-dimensional geometry}, we concentrate on the original mesh and the branch $\Phi_\mathrm{d}$, which will help us to gain some insights into mixing using, in particular, the observed elastic turbulent but also the weakly-chaotic flow state.}

\subsection{Mixing in viscoelastic fluids}
\label{subsec:mixing}

An interesting application of viscoelastic fluids is to utilise elastic turbulence at low Reynolds numbers for mixing solutes in a fluid or 
mixing of multiple fluids, which results in a homogenised fluid.
As a first approach we investigate the mixing of a passive scalar in the two-dimensional Taylor-Couette flow.

We start from the turbulent flow state, obtained at $\mathrm{Wi}=27.6$ after 250 rotations, without any scalar quantity present in the fluid. Then, we lower the Weissenberg number to the desired value and let the viscoelastic fluid flow equilibrate for another 100 rotations.
At this point, denoted time $t=0$, the scalar field is initialised with concentration $c_0$ on a ring with radius $r=0.432\,r_o$ and a width equal to the local grid spacing. 
{The initial scalar field is chosen close to the inner cylinder, where the radial velocity fluctuations, due to the increased hoop stress, are large.}
Afterwards, the fluid flow is evolved for an additional 400 to 500 rotations. The order parameter $\Phi_\mathrm{d}$ corresponding to these Weissenberg numbers is given by the blue circles in Fig.~\ref{fig:OPhysteresis}.

\begin{figure}
\centering
\includegraphics[width=0.45\textwidth]{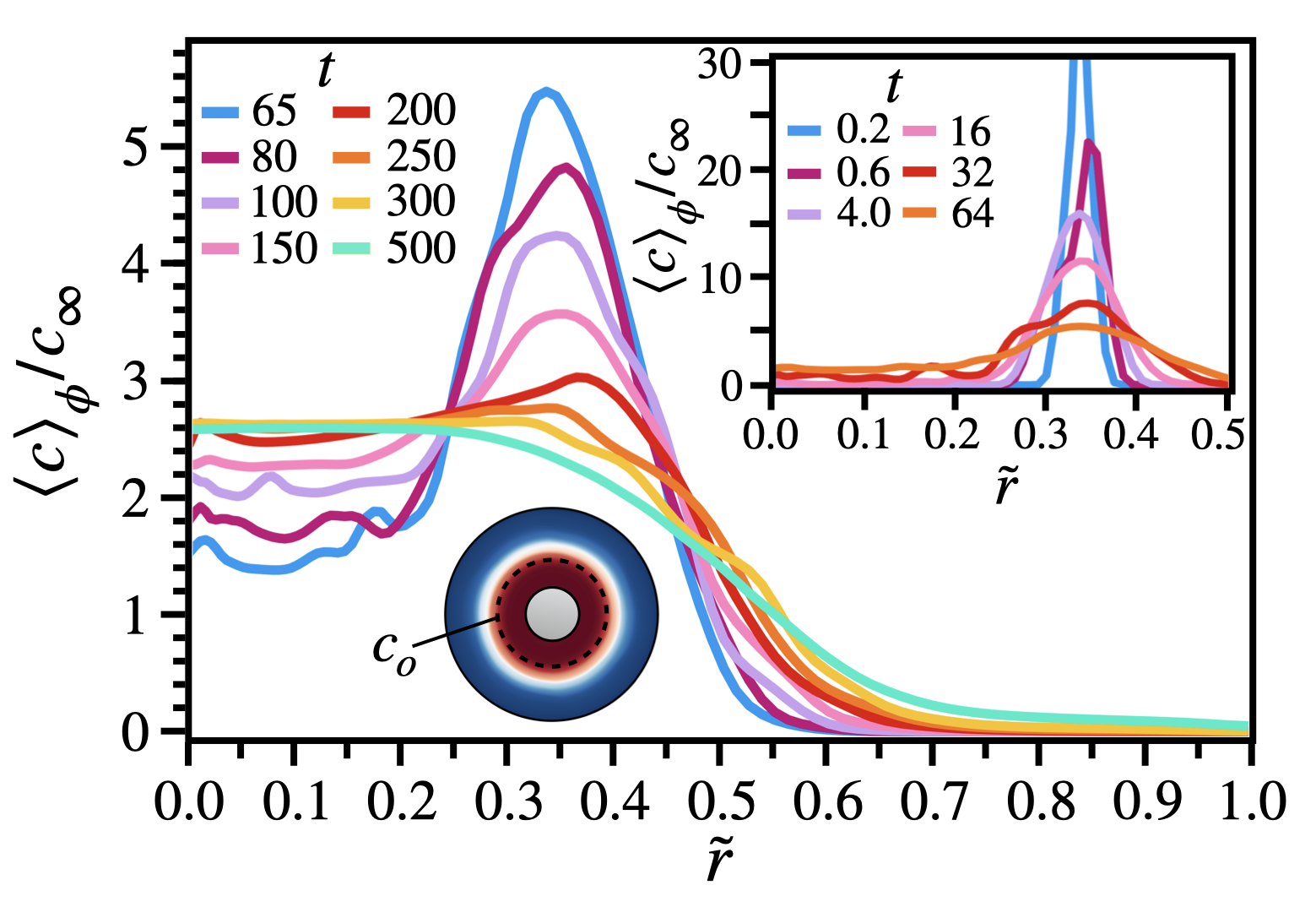}
\caption{
Averaged normalised concentration field of the passive scalar $\langle c \rangle_\phi /c_\infty$ plotted versus the normalised radial position $\tilde r = (r-r_i)/(r_o-r_i)$, where the averaging is over the azimuthal direction $\phi$, $c_\infty$ is the concentration of the completely mixed state.
The profiles are shown for different times $t$ at $\mathrm{Wi} = 10.68$, {with early times in the top right inset. The bottom left inset displays a snapshot of the concentration field at $t=598$. 
{The dashed ring indicates the initial position of the scalar field.}
}}
\label{fig:mixingprofile}
\end{figure}

\begin{figure}
\centering
\includegraphics[width=1.0\columnwidth]{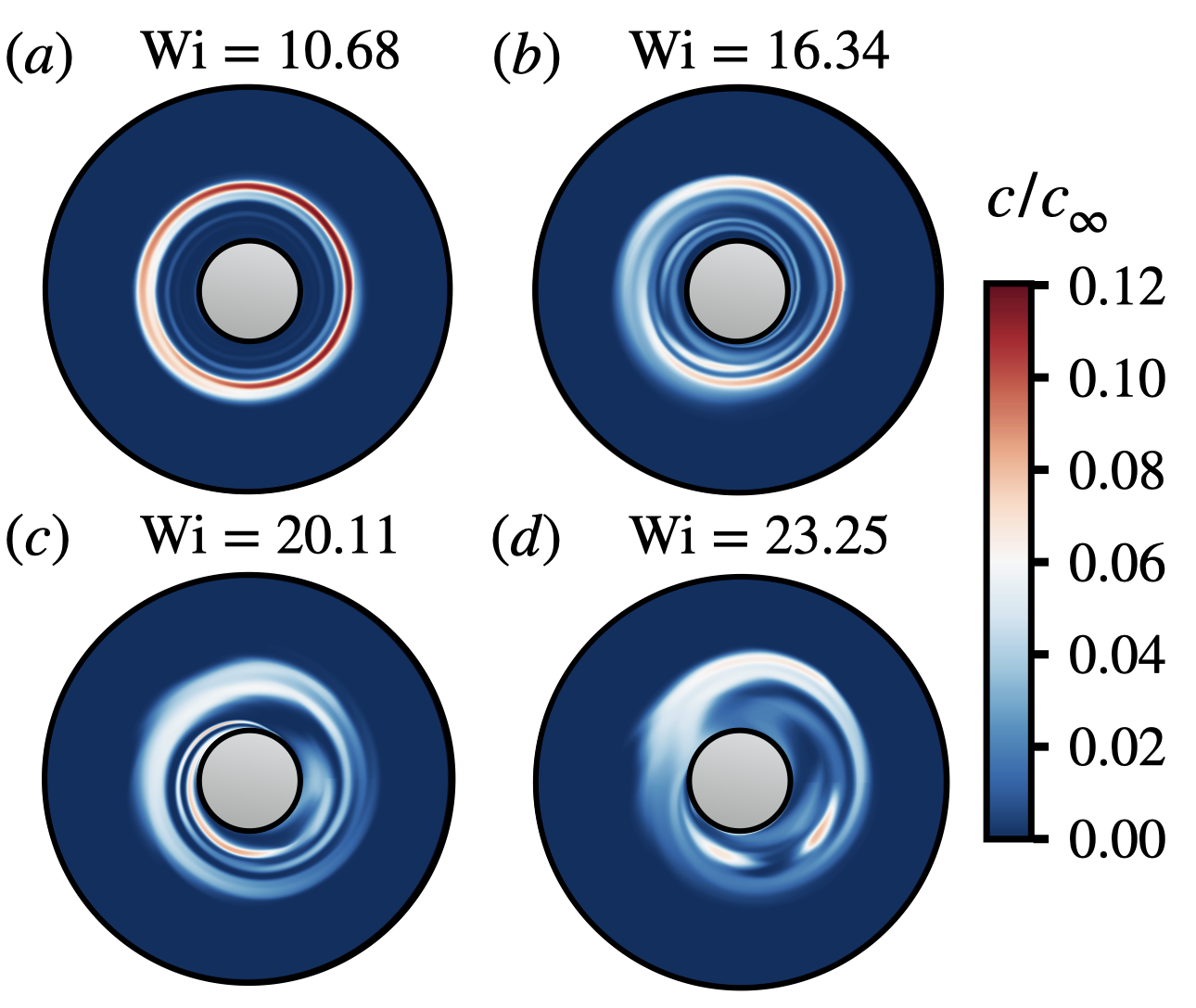}
\caption{
Snapshots of the normalised concentration field $c(r,\phi)/c_\infty$
for (a) $\mathrm{Wi} = 10.68$, (b) $\mathrm{Wi} = 16.34$, (c) $\mathrm{Wi} = 20.11$ and (d) $\mathrm{Wi} = 23.25$ at time $t=16$.
}
\label{fig:mixingSnap}
\end{figure}

In Fig.\ \ref{fig:mixingprofile} we plot the azimuthally averaged concentration field $\langle c(r,t) \rangle_\phi$ versus the normalised radial position \mbox{$\tilde r = (r-r_i)/(r_o-r_i)$} at different times in the elastic turbulent regime at $\mathrm{Wi}=10.68$, where
\begin{equation}
\langle c(r,t) \rangle_\phi = \frac{1}{2\pi} \int_0^{2\pi} c(\mathbf{r},t) d\phi \, .
\end{equation}
The concentration field is normalised by the concentration of a uniformly distributed solute, \mbox{$c_\infty = A^{-1} \int_A c(\mathbf{r},t) dA$}, {where $A$ is the area between the two cylinders}.
Initially, the strongly peaked distribution spreads quickly along the radial direction towards the inner cylinder (see inset of Fig.~\ref{fig:mixingprofile}). 
Moreover, for increasing \Wi we find that the concentration field spreads faster towards the inner cylinder because velocity fluctuations become stronger.
This is illustrated in Fig. \ref{fig:mixingSnap}, where we plot snapshots of the concentration field at time $t=16$ for (a) $\mathrm{Wi} = 10.68$, (b) $\mathrm{Wi} = 16.34$, (c) $\mathrm{Wi} = 20.11$ and (d) $\mathrm{Wi} = 23.25$. The figure shows how the concentration field is folded inwards by radial flow, which results in a spiral pattern of the concentration field. Random fluctuations in the radial velocity field lead to significantly different spreading of $c$ at different \Wi. 
Overall, the figure clearly demonstrates the initial rapid and random mixing for increasing \Wi.
With increasing time the concentration field becomes more evenly distributed in the region near the inner cylinder and a step-like profile develops (see Fig.~\ref{fig:mixingprofile}). Here, mixing slows down and at long times the concentration field develops towards a uniform distribution where $c = c_\infty$.
A snapshot of the concentration field for {$\mathrm{Wi} = 10.68$ at $t=598$} is shown in the bottom inset of Fig. \ref{fig:mixingprofile}.

\subsubsection{Characterising mixing}

We characterise the degree of mixing by the standard deviation of the concentration field normalised by its initial value,
\begin{equation}
\Delta c \equiv \sqrt{
\frac{\langle [ c(\mathbf{r},t)  -  c_\infty ]^2  \rangle }{\langle [ c(\mathbf{r},0)  - c_\infty ]^2  \rangle } } \, ,
\end{equation}
where $\langle g \rangle =  {A}^{-1} \int_A  g \, dA$ indicates an average of $g$ over the area $A$ between the two cylinders. The normalised 
{standard deviation}
$\Delta c$ is such that initially $\Delta c = 1$ and $\Delta c= 0$ for the completely mixed state with $c=c_\infty$.

\begin{figure}
    \centering
    \includegraphics[width=.45\textwidth]{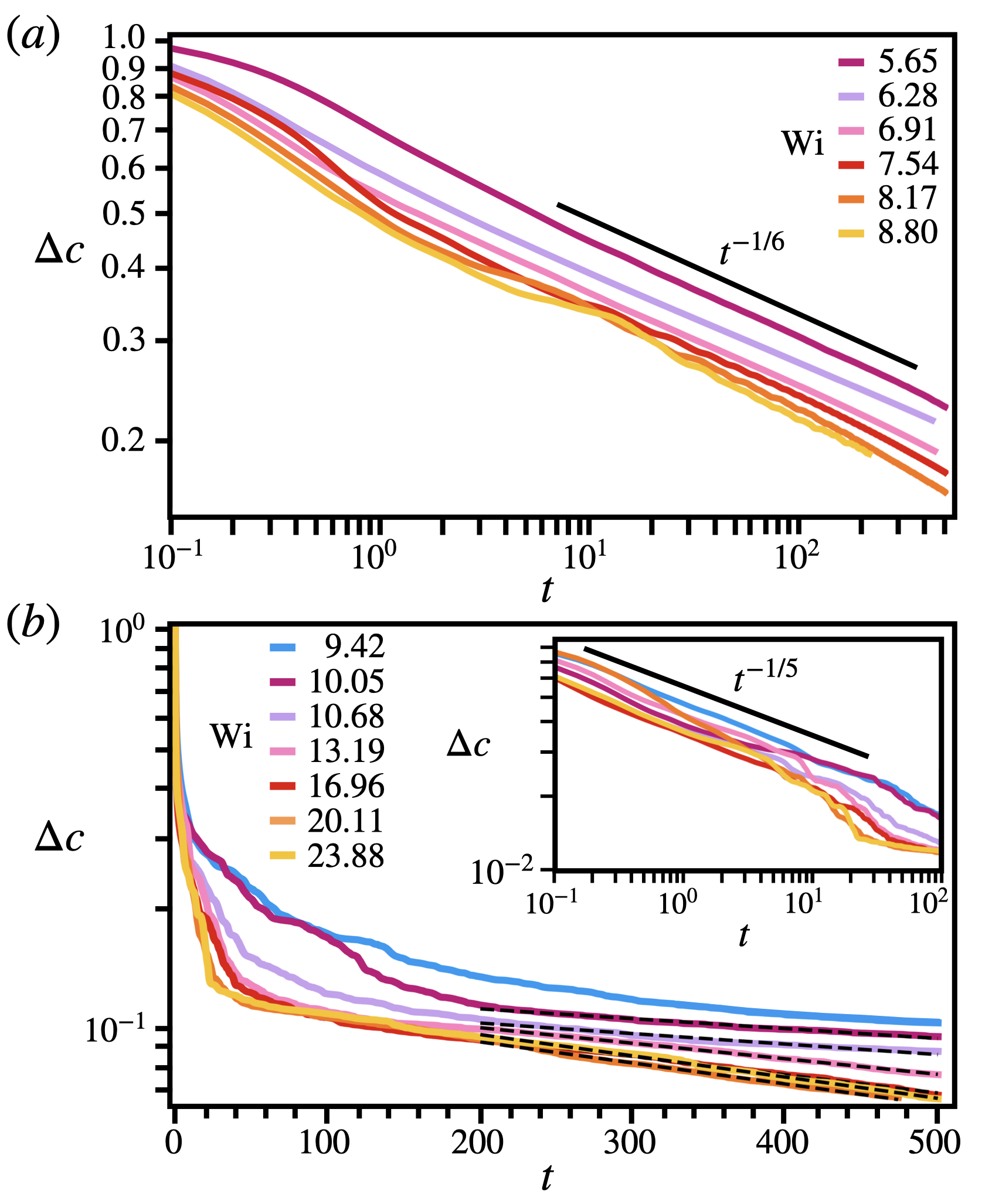}  
    \caption{
    Normalised 
    {standard deviation}
    of the concentration field $\Delta c$ plotted versus time $t$ at (a) Weissenberg numbers \mbox{$5.65 
    {\leq}
    \mathrm{Wi}\leq 8.80$} corresponding to the weakly-chaotic state and at (b) Weissenberg numbers $\mathrm{Wi}\geq {9.42}$, in the elastic turbulent state. In (a) power-law scaling is observed with an exponent close to $-1/6$. In (b) fits with the exponential function $\exp(-\alpha_\mathrm{mix} t)$ are indicated by the dashed lines, where $\alpha_\mathrm{mix}$ denotes the mixing rate. 
{Inset: For initial times a power-law decay is observed with an exponent close to $1/5$, which then enters an exponential decay.}
}
\label{fig:mixlowhigh}
\end{figure}

{In the laminar case, $\mathrm{Wi} < 5.03$, we observe a small time-independent radial flow field, which we think
to be artificial. In this regime, mixing is completely governed by diffusion, which in the present work is neglected ($D=0$). 
Thus, to analyse mixing and see the effect of irregular flow fields,
we focus on cases where  $\mathrm{Wi} > 5.03$.}

In the regime of weakly-chaotic flow, velocity fluctuations are increased, the velocity spectrum displays pronounced peaks and no 
power-law scaling is observed, hence, %although 
elastic turbulence does not occur (see Fig.~\ref{fig:OPhysteresis}). In Fig.~\ref{fig:mixlowhigh} (a) we plot the standard deviation of the concentration field $\Delta c$ versus time $t$ for Weissenberg numbers $5.03 < \mathrm{Wi} \leq 8.17$, \textit{i.e.}, in the weakly-chaotic flow state.
Initially, the standard deviation decreases stronger at larger \Wi, since the fluctuations in the flow field are larger. Then, for 
all values of $\mathrm{Wi}$ the standard deviation enters power-law scaling with an exponent close to $-1/6$.
{In contrast to what is expected for smooth chaotic flow, we do not find an exponential decay for $\Delta c$. However, this is in line with
reports in Ref.\ \cite{aref2017frontiers} that close to non-moving no-slip walls, power-law scaling is expected. Still we can conclude that}
the enhanced fluctuations in the velocity field increase mixing within the fluid.

{
In Fig.~\ref{fig:mixlowhigh} (b) we plot $\Delta c$ versus time $t$ for Weissenberg numbers $\mathrm{Wi} \geq 9.42$ in the elastic 
turbulent state. We observe an initial rapid spreading of the concentration field due to the broadening of the initial concentration 
ring as illustrated in the inset of Fig.\ \ref{fig:mixingprofile}.  As the inset of  Fig.~\ref{fig:mixlowhigh} (b) shows, the spreading first follows 
a transient power law, which we attribute again to the closeness of the non-moving inner cylinder. Ultimately, the spreading crosses over 
to an exponential decay, when the concentration has developed the step-like distribution, which then broadens further 
(see Fig.\ \ref{fig:mixingprofile}). Note, the concentration spreads preferentially inwards due to the cylindrical geometry and the close 
proximity of the initial profile to the inner boundary cylinder. Initially, the concentration profile broadens symmetrically in the radial direction, 
as the top right inset of Fig.\ \ref{fig:mixingprofile} shows. However, at smaller radii the available area is smaller and thus the concentration 
increases more strongly. Furthermore, once the inner boundary is reached, the concentration builds up.
}

We can fit the 
{the long-time tails} by an exponential decay, $\Delta c \sim \exp(-\alpha_\mathrm{mix} t)$, where $\alpha_\mathrm{mix}$ is the mixing rate. The fits are indicated {in the plot}
with the black dashed lines, which start at $t=200$ for all \Wi. 
We have chosen the 
starting time 
{of the fits} such that all slopes of the standard deviation in the elastic turbulent regime can be well fitted by an exponential tail. Moreover, increasing the fitting range for larger Wi does not alter the results.

{We summarise the situation so far, for the case when 
initiating a strongly peaked radial distribution of the concentration field.
In the weakly-chaotic flow state, we observe a power-law decay of the standard deviation $\Delta c$ from the uniform distribution,
which deviates from the expected exponential behaviour. We attribute this to the presence of the resting inner cylinder \cite{aref2017frontiers}.
In the elastic turbulent state, a transient power-law scaling of $\Delta c$ occurs, again attributed to the bounding inner wall, which crosses 
over to exponential decay, once the concentration assumes a step-like distribution. This raises the question, how does the concentration 
evolve when we start from an initial step distribution? In this case, we again recover exponential decay in the elastic-turbulent state, while
we now also observe it for the weakly-chaotic state, as Figs.\ \ref{fig:profilestep} in Appendix\ \ref{secA1} show. 
Thus, for the strongly peaked distribution in the weakly-chaotic flow, we also expect exponential decay to occur in the long-time limit, 
after the step-like distribution has developed. However, this is beyond feasible simulation times.}

\begin{figure}
    \centering
    \includegraphics[width=.475\textwidth]{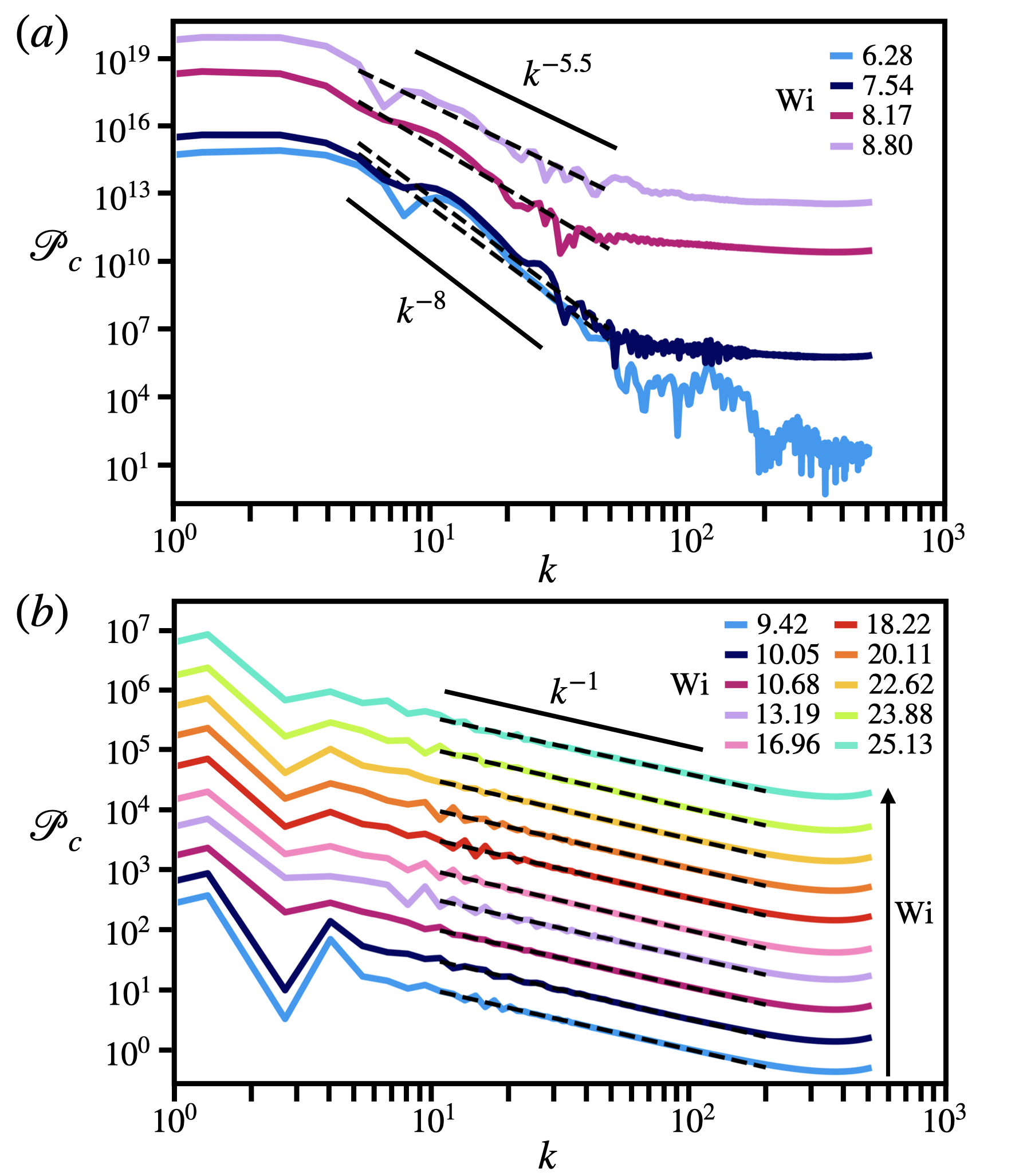}  
    \caption{
    Radial power spectral density ${\mathcal{P}_c} (k)$ of the scalar concentration field as a 
    function of the radial wave number $k$ for different \Wi at time $t=550$.
    The data in (a) and (b) are arbitrarily offset by half a decade  for clarity. 
    {(a) Weakly-chaotic state, (b) elastic turbulent state.}
    Power laws $k^{-\alpha}$ are fitted and plotted as black dashed lines
    {in the range 
    $10 \leq k \leq 200$ 
    in (a) and}
    in the range $10 \leq k \leq 100$ in 
    {(b).}
}
\label{fig:scalarPSD}
\end{figure}

\subsubsection{Batchelor regime of mixing}

The exponential scaling observed in the standard deviation suggests mixing within the fluid occurs in the Batchelor regime. 
To further investigate this, we compute the radial power spectral density of the concentration field following Ref.\ \cite{pierrehumbert2000batchelor}.
We start by introducing the Fourier transform in polar coordinates: $\hat c(\mathbf{k}) = \mathcal{F}\{c\}(\mathbf{k})$, 
where the wave vector $\mathbf{k}$ is represented by the radial wave number $k$ and the polar angle $\theta$.
In order to perform the Fourier transform in the radial direction, which requires a regular radial spacing of the grid, the concentration field is approximated by a cubic spline fit with constant spacing. Then, averaging over the azimuthal direction due to the cylindrical symmetry 
{for a specific time $t$} 
the radial power spectrum of the concentration field is obtained as \cite{pierrehumbert2000batchelor}
\begin{align}
    \mathcal{P}_c (k)  = \frac{1}{(2\pi)^2} \int_0^{2\pi}  \lvert \hat c(\mathbf{k}) \rvert^2 k d\theta  \, ,
\end{align}
The power spectrum $\mathcal{P}_c (k)$ is plotted in Fig.\ \ref{fig:scalarPSD}
{for the weakly-chaotic (a) and the elastic turbulent (b) state.}
{Note that
{in (b) the form of  $\mathcal{P}_c (k)$}
does not depend on the choice of the time $t$, as long as the step-like distribution in the elastic turbulent flow 
has been fully developed. Thus, we}
observe a Batchelor spectrum with power scaling $k^{-1}$ in the elastic turbulent regime, $\mathrm{Wi}\geq 9.42$, which is also 
{measured}
in experiments \cite{groisman2001efficient}.
{In contrast, in the weakly-chaotic state we obtain a power scaling with exponent $-5$ or even larger, which we attribute to the power-law scaling
of $\Delta c$.}

\subsubsection{Mixing rate}

\begin{figure}
    \centering
         \includegraphics[width=.45\textwidth]{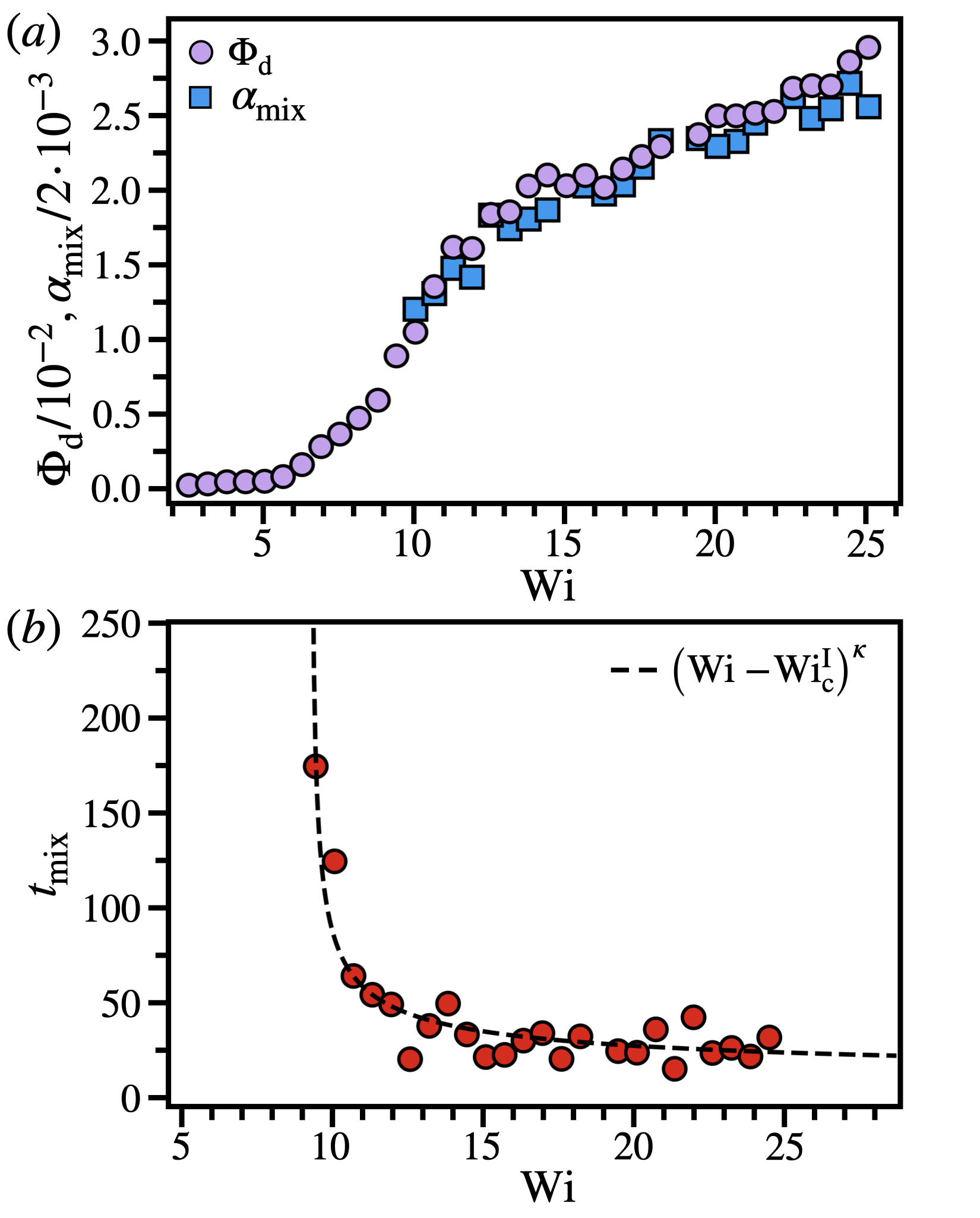}  
\caption{(a) Order parameter $\Phi_\mathrm{d}$ and the mixing rate $\alpha_\mathrm{mix}$ versus the Weissenberg number \Wi. (b) Mixing time $t_\mathrm{m}$ versus the Weissenberg number \Wi. The dashed line indicates the scaling law $(\mathrm{Wi} -\mathrm{Wi_c^I})^{\kappa}$, with $\mathrm{Wi_c^I}=9.26 \pm 0.04$ and $\kappa = -0.52 \pm 0.03$.
}
\label{fig:mixorderparam}
\end{figure}

Now, we take a closer look at the mixing rate and how it relates to the order parameter.  In Fig.\ \ref{fig:mixorderparam} (a) we plot both the order parameter $\Phi_\mathrm{d}$, obtained before in Fig.~\ref{fig:OPhysteresis}, and the mixing rate $\alpha_\mathrm{mix}$, obtained from the fits in Fig.~\ref{fig:mixlowhigh}\ (b).
We observe a strong correlation between $\alpha_\mathrm{mix}$ and $\Phi_\mathrm{d}$, which indicates that increased velocity fluctuations increase the mixing within the fluid. 
Since mixing below $\mathrm{Wi} < 9$ follows algebraic scaling, a mixing rate cannot be determined and is not included in the plot.
We conclude that our order parameter is therefore also a good measure for mixing within the fluid.

Moreover, we define a mixing time from the standard deviation of the concentration field.  We determine the fluid to be well-mixed when the standard deviation is small, $\Delta c  \approx 0.14$, and define the mixing time $t_\mathrm{mix}$, when this value is reached. It was chosen such that $\Delta c$ in the elastic turbulent state falls below 0.14 within the total simulation time, while this value is not reached in the weakly-chaotic state occurring below $\mathrm{Wi} < 9.4$.
In Fig.~\ref{fig:mixorderparam} (b) we plot the mixing time versus \Wi. The figure shows a strong decrease in $t_\mathrm{mix}$ with increasing \Wi. 
The mixing time in the turbulent regime at $\mathrm{Wi} \geq 9.4$ roughly follows the scaling $t_\mathrm{mix} \sim \left( \mathrm{Wi} - \mathrm{Wi_c^I} \right)^{-1/2}$ with $\mathrm{Wi_c^I} = {9.26}$.
The mixing time includes both the 
initial mixing and the 
{exponential} mixing afterwards. It is sensitive to fluctuations during the initial 
mixing, where random velocity fluctuations in the radial directions have a large impact on the overall mixing time. Thus, it exhibits large variations.
Nevertheless, 
a strong decrease in the mixing time is observed in the turbulent flow state, which demonstrates the impact of elastic turbulence on mixing.

\subsubsection{Mixing regimes}

\begin{figure}
    \centering
    \vspace*{-2pt}
    \includegraphics[width=.45\textwidth]{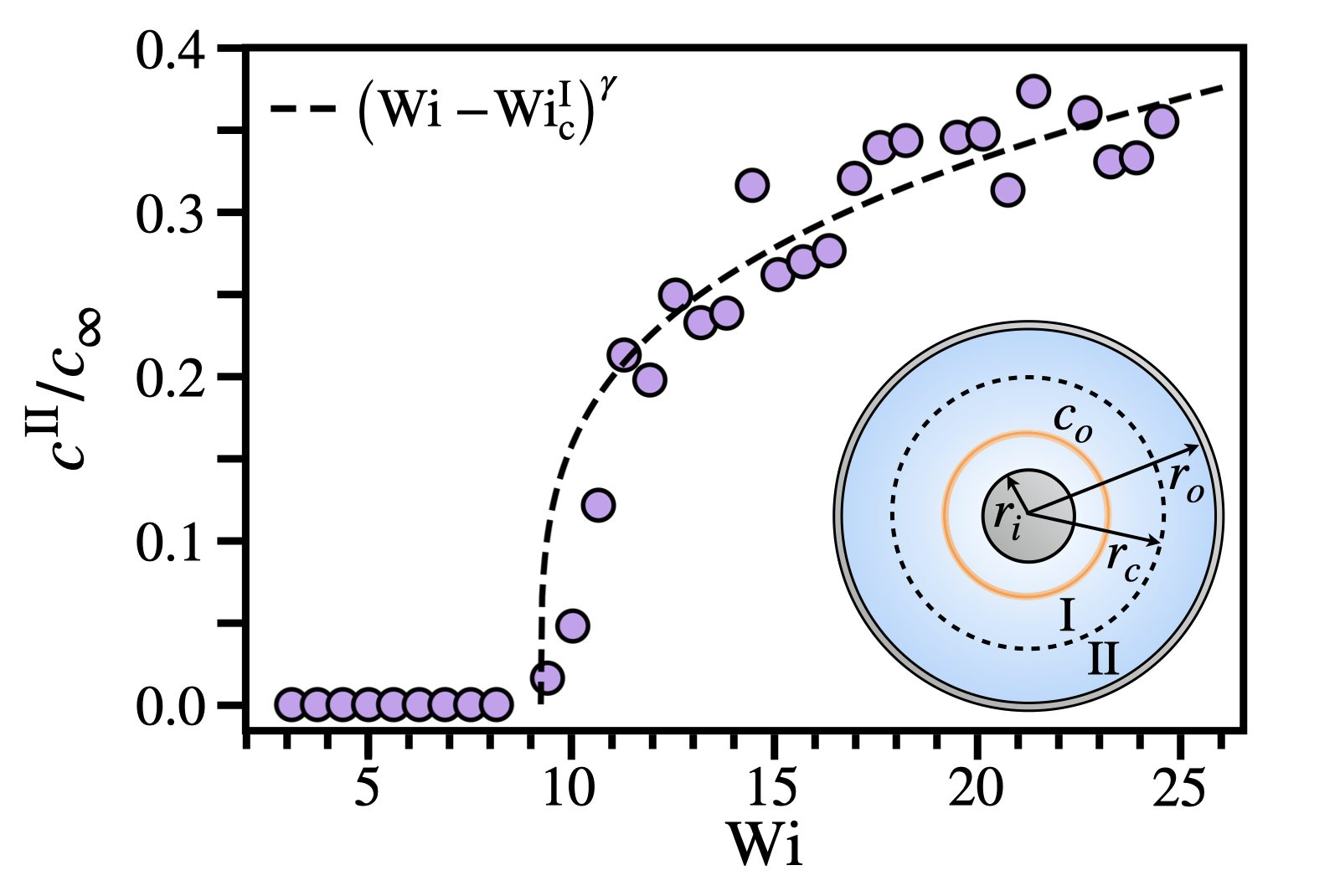}  
\caption{
Normalised concentration $c/c_\infty$ in region II (see inset)
averaged between times $t=400$ and $t=500$. The dashed line shows the scaling law $(\mathrm{Wi} -\mathrm{Wi_c})^{\gamma}$ with $\mathrm{Wi_c^{I}}=9.26$ and $\gamma = 0.28 \pm 0.03$.
}
\label{fig:mixregion}
\end{figure}

Lastly, we determine the degree of mixing by examining the concentration that develops in a specific region. To this end, we divide the Taylor-Couette geometry into two equal areas: the inner region (region $\mathrm{I}$), near the inner cylinder with $r_i < r < r_c$ and the outer region (region $\mathrm{II}$), near the outer cylinder with $r_c < r < r_o$. The dividing radius $r_c$ is determined from the condition of equal areas and thus is given by $r_c = \sqrt{17/32} {\,r_o} \approx 0.73 {\,r_0}$. 
Initially, the passive scalar is located in the inner region.
Therefore, we consider the developing concentration in the outer region $\mathrm{II}$ as a good indicator for the degree of mixing.
In Fig.~\ref{fig:mixregion} we plot the normalised concentration $c^\mathrm{II}/c_\infty$ in \mbox{region $\mathrm{II}$} as a function of the Weissenberg number. The concentration is averaged 
{over region II and}
between times $t=400$ and $t=500$, {which is chosen such that it is large enough to give an accurate average value, but not too large such that the concentration varies significantly (see Fig. \ref{fig:mixlowhigh})}. We again observe scaling behaviour, where $c^\mathrm{II} \sim (\mathrm{Wi} -\mathrm{Wi^{I}_  c})^{\gamma}$ with $\gamma= 0.28 \pm 0.03$ and $\mathrm{Wi_c^I}=9.26$.
Comparing with the scaling we found for $t_\mathrm{mix}$, we thus obtain the scaling relation $c^{\,\mathrm{II}} \sim t_\mathrm{mix}^{-1/2}$.
We checked that his scaling relation does not depend on varying the initial time of the averaging time interval  between 250 and 450.
{However, we expect the observed scaling exponent to depend on the initial radial position of the peaked distribution.}

\section{Conclusion}
\label{sec:conclusion}
We have investigated how advection by the flow field of elastic turbulence determines mixing within a viscoelastic fluid in a two-dimensional Taylor-Couette flow.
For this we used numerical solutions of the Oldroyd-B model with the program \OpenFOAM.
The Weissenberg number is varied while keeping the Reynolds number very small and constant.
In contrast to our earlier work \cite{vanBuel2018elastic}, we identified the transition towards elastic turbulence as a subcritical transition. 
When lowering the Weissenberg number, elastic turbulence continues to exist below the critical Weissenberg number $\mathrm{Wi}_c = 9.95$, determined earlier for increasing \Wi \cite{vanBuel2018elastic}. 
Then, below $\mathrm{Wi_c^I} = 9.26$ a weakly-chaotic flow occurs, which ultimately becomes laminar for $\mathrm{Wi} \leq 5.03$.

The turbulent and weakly-chaotic secondary flows have non-zero radial components, which is ideal for mixing at the micrometer scale. 
{Chaotic advection} in the weakly-chaotic flow state causes mixing, where the standard deviation of the concentration field shows power-law scaling.
The elastic turbulent state displays % 
initial {power-law} mixing followed by an %
exponential decay in the standard deviation.
The exponential decay corresponds to the regime of Batchelor mixing, also observed in experiments on mixing in elastic turbulence \cite{groisman2001efficient}.
The exponential decay or the mixing rate coincides with the secondary-flow order parameter. Finally, for the elastic turbulent state we also introduced a mixing time and mixing efficiency, which follow scaling laws {for our choice of the initial concentration distribution}.

Our numerical work confirms that elastic turbulence can be harnessed for mixing a passive scalar in viscoelastic fluids at 
very low Reynolds numbers {and the Batchelor regime is identified, both results are also observed in experiments \cite{groisman2001efficient, groisman2004elastic}}. Moreover, we have demonstrated that the secondary-flow order parameter is strongly correlated 
to the mixing rate and hence it is also a good indication for the degree of flow-induced mixing.
{An important question for future research is how the presence of elastic waves, which have recently been observed in simulations 
in the Taylor-Couette geometry by Song et al. \cite{song2023self}, impacts the overall mixing efficiency, the mixing rate and the mixing time.}

\backmatter

%\bmhead{Supplementary information}

\bmhead{Acknowledgments}

We acknowledge support from the Deutsche For\-schungs\-ge\-mein\-schaft in the framework of the Collaborative Research Center SFB 910
{and from Technische Universit\"at Berlin}.

%Some journals require declarations to be submitted in a standardised format. Please check the Instructions for Authors of the journal to which you are submitting to 
%see if you need to complete this section. If yes, your manuscript must contain the following sections under the heading `Declarations':

\section*{Author contribution statement}
All authors contributed equally to this work.
%RB performed the simulations. Both authors were equally involved in the analysis of the results and writing the manuscript.

\subsection*{Declarations}

\subsubsection*{Conflict of Interest}
The authors have no conflicts to disclose.

\subsubsection*{Data Availability Statement}
The data that support the findings of this study are available from the corresponding author upon reasonable request.

\begin{appendices}

\section{ }
\subsection*{{Initial step distribution of the concentration field}}\label{secA1}

{
Here, we investigate the effect of the initial distribution of $c$. In Sec. \ref{subsec:mixing}, we found that 
{the standard deviation of $c$ in the elastic turbulent state displays exponential scaling once a step-like distribution
has developed. To investigate this further, we performed simulations where we directly}
start from a step distribution,
{also in the weakly chaotic state,}
and analyse the standard deviation of $c$. Other than the initial distribution of $c$, we employ the same conditions as described in Sec.~\ref{subsec:methods}. }

\begin{figure}
    \centering
    \includegraphics[width=.475\textwidth]{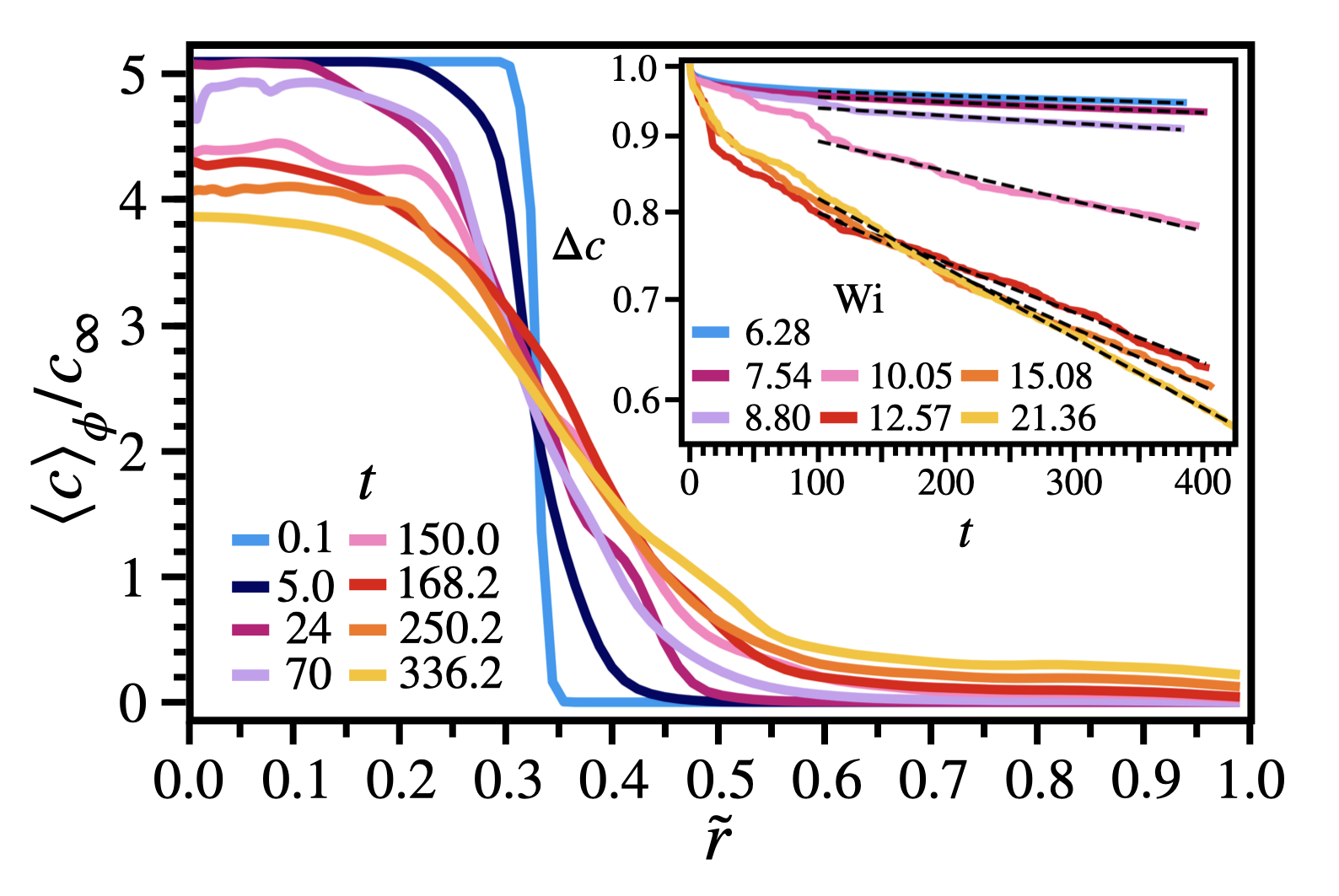}  
    \caption{{
Averaged normalised concentration field of the passive scalar $\langle c \rangle_\phi /c_\infty$ plotted versus the normalised radial position $\tilde r = (r-r_i)/(r_o-r_i)$, where the averaging is over the azimuthal direction $\phi$, $c_\infty$ is the concentration of the completely mixed state and the initial profile is a step distribution. The profiles are shown for different times $t$ at $\mathrm{Wi} = 21.36$. The inset displays the normalised standard deviation of the concentration field $\Delta c$ plotted versus time $t$ at Weissenberg numbers $6.28\geq \mathrm{Wi}\geq {21.36}$. Fits with the exponential function $\exp(-\alpha_\mathrm{mix} t)$ are indicated by the dashed lines, where $\alpha_\mathrm{mix}$ denotes the mixing rate. 
}}
\label{fig:profilestep}
\end{figure}

{
First, we plot the azimuthally averaged concentration $c$ as a function of the normalised radial position $\tilde r = (r-r_i)/(r_o-r_i)$, starting from a step distribution, for $\mathrm{Wi}=21.36$ in Fig. \ref{fig:profilestep}. In time, the concentration decreases near the inner cylinder and increases towards the outer cylinder, similar to what we observed in Fig. \ref{fig:mixingprofile}. Second, the standard deviation $\Delta c$
{versus time $t$}
of this case is plotted in the inset of Fig. \ref{fig:profilestep} 
{together with results for other $\mathrm{Wi}$ including the weakly-chaotic state.}
We observe exponential scaling for $t>100$, for all $\mathrm{Wi}\geq 6.28$.
Thus, in contrast to the results in Sec. \ref{subsec:mixing}, we now observe exponential scaling of $\Delta c$ 
{even}
in the weakly-chaotic
{state while for the elastic turbulent state it starts close to $\Delta c =1$.}
Therefore, we conclude that the power-law scaling is related to the ring-like distribution and the exponential scaling is related to the step-like distribution.
}

\end{appendices}


\bibliography{literature}


\end{document}